\documentclass[preprint,12pt]{elsarticle}

 \usepackage{graphics}
 \usepackage{epsfig}

\usepackage{amssymb}

\usepackage[dvips]{color}
\usepackage{url}
\usepackage{subfigure}

\begin{document}

\begin{frontmatter}




\title{A near real-time framework for extracting tip-sample forces in dynamic atomic force microscopy (dAFM)}
\author{David Busch, Qingze Zou}
\author{Baskar Ganapathysubramanian \corref{cor1}}
\address{Department of Mechanical Engineering,
    2100 Black Engineering, Iowa State University, Ames, IA
    50010,USA}
\cortext[cor1]{Corresponding author: B. Ganapathysubramanian, Fax: 515 294-3261,  Email: baskarg@iastate.edu, URL: http://www3.me.iastate.edu/bglab/}

\begin{abstract}
The atomic force microscope (AFM) is a versatile, high-resolution tool used to characterize the topography and material properties of a large variety of specimens at nano-scale. The interaction of the micro-cantilever tip with the specimen causes cantilever deflections that are measured by an optical sensing mechanism and subsequently utilized to construct the sample topography. Recent years have seen increased interest in using the AFM to characterize soft specimens like gels and live cells. This remains challenging due to the complex and competing nature of tip-sample interaction forces -- large tip-sample interaction force is necessary to achieve good signal-to-noise ratios; However, large force tends to deform and destroy soft samples. In situ estimation of the local tip-sample interaction force is needed to control the AFM cantilever motion and prevent destruction of soft samples while maintaining a good signal-to-noise ratio. This necessitates the ability to rapidly estimate the tip-sample forces from the cantilever deflection during operation.

This paper proposes a first approach to a near real-time framework for tip-sample force inversion. We pose the inverse problem of extracting the tip-sample force as an unconstrained optimization problem. A fast, parallel forward solver is developed by utilizing graphical processing units (GPU). This forward solver shows an effective 30000 fold speed-up over a comparable CPU implementation, resulting in milli-second calculation times. The forward solver is coupled with a GPU based particle-swarm optimization implementation. We illustrate the framework on three classes of tip-sample interaction inversions. Each of these inversions is performed in sub-second timings, showing potential for integration with on-line AFM imaging and material characterization.

\end{abstract}


\end{frontmatter}


\section{Introduction}
The ability to characterize soft materials on the micro/nano-scale has significant implications to several areas in science ranging from fundamental studies in polymer physics \cite{Loos_AFMReview_05,Beekmans2004893,hahm:4730} to applied bio-engineering~\cite{Jena_CellB_02,LinH99}, where understanding nanoscale behavior and evolution is essential. 

Dynamic AFM imaging \cite{Hansma1994} is a very effective technique to interrogate surface topography of soft samples \cite{Loos_AFMReview_05,Parot2007}, particularly for live biological samples in their physiologically friendly liquid environment \cite{Miles_Sci_03,horton_00}. \footnote{For instance, by using dynamic AFM imaging, time evolving phenomena like crystallization of polymers \cite{Beekmans2004893} and the dehydration process of collagen \cite{LinH99,Feninat2001} have been experimentally revealed for the first time.} Dynamic atomic force microscopy (dAFM) or intermittent contact mode AFM utilizes a micro-cantilever fixed-free beam to interrogate samples. The cantilever base is driven by a piezo-actuator to oscillate, causing the free tip to tap (i.e., come into intermittent contact with the sample). The oscillation amplitude and phase with respect to the cantilever base are measured and the amplitude is maintained around a set-point value via feedback control. The measured phase and amplitude data are then utilized to construct the sample topography and also related to the material properties of the sample \cite{cleveland1998energy, Garcia_TapForceSize_99}. 

\begin{figure} [ht]
\centering
\includegraphics[scale=0.2]{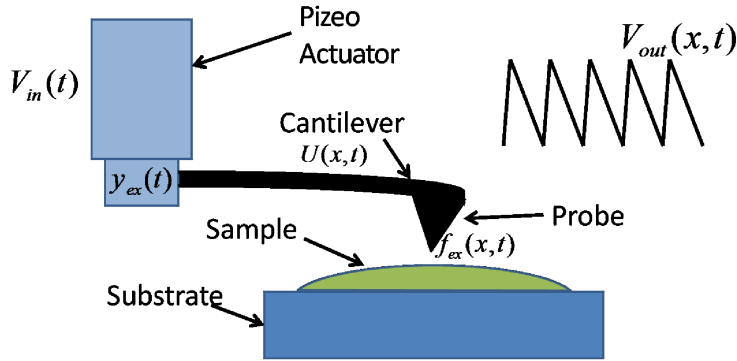}
\caption{AFM diagram}
\label{AFM_Diagram}
\end{figure} 

Although by using dynamic-mode imaging, the detrimental sliding force on the sample has been largely reduced, the applied normal (tapping) force can still be large and result in not only imaging distortion, but more seriously, sample deformation and damage that can completely modify the sample \cite{Parot2007}. Large normal force, however, is needed in dynamic-mode AFM imaging to ensure imaging quality (i.e., high signal to noise ratio).  The requirement of rapid scanning (high-speed) imaging of specimens further exacerbates these problems~\cite{Raman2006,Sader2007}. {\it The challenge in tackling these hurdles lies in the need to maintain a small tip-sample interaction force during the scanning process}. Therefore, as a key first step to tackle this challenge, estimating the tip-sample interaction force -- accurately and in real-time -- is essential.

Current methods for tip-sample force inversion generally require significant post-processing time and are, thus, incapable of addressing sample deformation and destruction in real-time~\cite{Raman08}.  Off-line inverse problems have been formulated to estimate tip-sample interaction forces using conjugate gradient optimization \cite{Chang2005} with limited success. The availability of newer computing methods, such as general purpose graphical processing unit computing, opens up the possibility of near real-time inversion. This paper focuses on formulating and implementing a parallel computational framework for fast inversion of tip-sample forces by using the hardware and software capabilities of GPU's and Compute Unified Device Architecture (CUDA), respectively. To the authors best knowledge, this is the first time that near real-time (sub-second) inversion of tip-sample forces has been showcased. Other contributions include: 1) formulating the problem of estimating the tip-sample interaction force as an inverse problem posed as an unconstrained optimization problem; 2) developing an ultra-fast predictive model for AFM dynamics based on parallel algorithms implemented on GPUs; 3)  applying gradient-free optimization techniques to quickly find a solution to the optimization problem; and 4) showcasing a hierarchy of models for inversion.

In Section.~2 we formulate the problem definition and pose the direct and inverse problems. Section.~3 details the computational developments and algorithms along with some timing and complexity studies. In Section.~4 we showcase the fast inversion framework on several examples of increasing complexity. 
\section{Problem formulation}
\subsection{Introduction to the problem}
\begin{figure} [ht]
\centering
\includegraphics[scale=0.5]{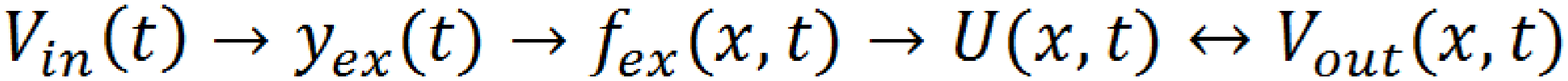}
\caption{Dynamic atomic force microscope force interaction process.}
\label{AFM_Eq}
\end{figure}

The physics of the sensing process demonstrates the difficulty in extracting the tip-sample interactions from the measured deflection data.  A schematic of the AFM sensing process is shown in Fig. \ref{AFM_Eq}.  

An AC input voltage ($V_{in}(t)$), usually a sinusoidal wave is sent to a piezoelectric actuator attached to the base of the cantilever (see Fig.~\ref{AFM_Diagram}), resulting in the oscillation of the base of the cantilever, $y_{ex}(t)$.  Then as the vibrating tip is brought into intermittent contact with the sample, the tip-sample interaction force ($f_{ex}(x,t)$) is induced, which in-turn, results in the change of the oscillation (or vibration) pattern at the cantilever tip, $U(x)$.  The tip deflection is measured using an optical sensing scheme.  

We approach the problem of extracting the tip-sample interaction force $f_{ex}(x,t)$ in two stages-- first solve the forward dynamics problem of calculating cantilever deflections given a tip-sample interaction force and subsequently solving associated inverse problem of calculating interaction forces given a cantilever deflection.

\subsection{The forward problem: Quantification of tip force effects on AFM cantilever}
The forward problem quantifies the relationship between the cantilever base displacement $y_{ex}(t)$ and the tip sample interaction $f_{ex}(x,t)$ with the tip deflection $U(x,t)$.  We model the cantilever dynamics by using Euler-Bernoulli (EB) beam theory.  The choice is driven by the following rational: (a) the EB model provides a more accurate description of AFM cantilever dynamics than the conventional simple harmonic oscillator model \cite{Rodriguez2002}; (b) from an inverse problem standpoint the EB model is computationally more tractable then finite element formulations \cite{Raman08,Rodriguez2002}, with minimal loss in fidelity~\cite{Sader2007,Resonant_Sader}; and (c) all assumptions made in EB beam theory is satisfied by an AFM cantilever \cite{Sader_JAP_98, Sader_JAP_00}.  

By the EB theory, the cantilever dynamics is modeled as \cite{Meirovitch2001}:
\begin{eqnarray}
EIu'''' (x,t) + 2\zeta \dot{u} (x,t) + \mu \ddot{u} (x,t) = f_{ex}(x,t) - \mu \ddot{y}_{ex} (t),
\label{EB_diff_eq}
\end{eqnarray}
where $u$ is the relative cantilever displacement with respect to the cantilever base; $E$, $I$, $\mu$ are the Young's modulus, moment of inertia, and mass per unit length, respectively; $\zeta$ is the viscous damping coefficient; $f_{ex}$ is the interaction force, and $y_{ex}$ is the cantilever base displacement.  It is a standard practice to convert this PDE into a set of ordinary differential equations (ODE) \cite{Meirovitch2001} by solving the free vibration eigenvalue problem and utilizing the orthogonality of the modal/eigen functions, $\Phi_i$.  The ODEs are solved for $\eta_i$ to compute the cantilever deflection $u(x,t) = \sum \Phi_i(x)\eta_i(t)$.  This results in a coupled set of ODEs for the coefficients of the modal functions,

\begin{eqnarray}
\ddot{\eta_i} (t) + 2 \zeta \omega_i \dot{\eta_i} (t) + \omega^2_i \eta_i (t) = F_i (t) - Y_i (t), 
\label{EB_ODE}
\end{eqnarray}
where $\omega_i$ are the corresponding modal resonance frequencies, $F_i (t) = \langle f_{ex}(x,t), \Phi_i (x) \rangle$ \footnote{$\langle A(x),B (x) \rangle$ is defined as the inner product of A and B over the length of the cantilever, $\langle A(x),B (x) \rangle = \int_0^L A(x)B(x) dx$}, and $Y_i (t) = \langle \mu \ddot{y}_{cal} (t), \Phi_i (x) \rangle$.  

Solving Eq.~(2) and hence the original beam dynamics equation~(1) to obtain the cantilever displacement U(x, t) requires that the tip-sample interaction force to be known. Tip-sample interactions are usually parametrized to account for different types of forces.  The simplest tip-sample interaction is an elastic response that can be modeled by Hooke's law \cite{Morris}:  \\
\begin{eqnarray}
f_{spring} = -k(u-h)
\label{spring}
\end{eqnarray}
where $u - h$ is the distance the cantilever tip has pressed into the sample (h is the datum), and $k$ is the local stiffness of the sample. More complex materials respond in a visco-elastic manner, dissipating some of the energy of the tip-sample interaction \cite{Raman08}.  This response is modeled using a spring-damper system given by:
\begin{eqnarray}
f_{spring-damper} = -k(u-h) - \zeta_s\dot{u}
\label{spring_damper}
\end{eqnarray}
where $\zeta_s$ is the viscous damping constant. More complex models can include Hertzian contact and Van der Waals forces~\cite{Raman08}.  

In summery, the definition of the forward problem is as follows: \newline \emph{{\bf FP:} Given the cantilever properties, ($E, I, \mu$), the cantilever base displacement $y_{ex}(t)$, and the parametrized tip-sample interaction force,(Eqns.~\ref{spring} or \ref{spring_damper}), calculate the cantilever deflection $U(x,t)$}.  

\subsection{The Inverse problem: Quantification of the tip-sample interaction force}
The inverse problem is defined as follows:  \newline \emph{{\bf IP:} Given the cantilever properties, ($E, I, \mu$), the measured cantilever tip deflection $U(x,t)$, and the given cantilever base displacement $y_{ex}{t}$, calculate the parametrized tip-sample interaction $f_{ex}(x,t)$.}  

One approach to solving the inverse problem is to convert it into an unconstrained optimization problem through the minimization of a chosen cost functional $\mathcal{J}$ that minimizes the difference between $U(x,t)$ and $u(x,t)$.  An appropriate choice of the cost functional $\mathcal{J}$ acts as a metric that quantifies the mismatch between a guess value of the tip-sample interaction and the true tip-sample interaction.  The choice of the cost functional plays a significant part in the accuracy and efficiency of the inversion process.  The proper choice of the cost functional ensures reasonable speed of calculation and a smooth phase space.  Extensive computational experiments suggested the use of the following cost functional:

\begin{eqnarray}
\mathcal{J}_{L2}^2 (F) = \int_0^{t_{max}} \left[ U(x,t) - u(x,t) \right]^2 dt
\label{j_l2}
\end{eqnarray}
where $U$ is the experimentally measured tip deflection.  $u$ is the calculated tip deflection for given the tip-sample interaction $F$ (i.e., by solving the forward problem {\bf FP}).  

The unconstrained optimization problem is posed as follows:  
\newline \emph{Given cantilever properties ($E, I, \mu$), cantilever base movement $y_{ex}(t)$, and the experimental cantilever tip deflection ($U(x,t)$), find the parametrized tip-sample interaction $F^*$ such that $\mathcal{J}_{L2} (F^*) \leq \mathcal{J}_{L2} (F)$ for any $F$, where $\mathcal{J}_{L2}$ is defined in Eqn.~\ref{j_l2}.}

\section{Computational framework for near real time inversion}
This section details the computational framework for solving the direct and inverse problems formulated in the previous section. A key challenge is the necessity of very fast force-inversion for real time dAFM imaging of soft samples to be possible. Posing the direct problem as a set of ODE's and the force inversion as an unconstrained optimization problem over these ODEs allows to leverage the computational advantages provided by GPUs. The rational for using GPU's is guided by the following reasons: (1) ability to construct a large set of forward problems in parallel; (2) faster analysis given faster memory accesses compared to CPUs; (3) GPU compute architecture is well suited for problems with minimal parallel dependencies; and (4) GPU compute architecture is well suited for problems where the computation-to-memory-access ratio is larger than one. A brief description of GPU hardware and CUDA software concepts utilized in the developed framework follows.

\subsection{CUDA and GPU computing}
Utilizing GPU's for computation is different than on CPU's.  GPU's require a large number of threads of execution that are processed in parallel to be efficient.  In contrast, CPU's are generally more efficient with few threads.  

\textbf{Memory:} GPUs (running CUDA) have very large computation capability compared to the speed at which they can access memory. GPU's hide this memory latency by performing computation and memory grabs simultaneously.  While sets of threads (called warps) are waiting for their data from memory, other warps get computed. The availability of hierarchies of memory allows significant room for designing algorithms to optimize memory access, thus enhancing speed.  We briefly describe the memory modes that are used in the current formulation; \emph{global, shared, texture, and constant}.
\begin{enumerate}
\item Global memory is the main memory storage on GPU and is the slowest to access requiring hundreds of clock cycles. Global memory is retrieved in groups of bytes for warps based on the requirements of the threads.  Warps can only grab memory that is in sequential order.\footnote{So if thread 0 needs memory from array position 0 and thread 1 requires array position 1000000, the warp will request two accesses to global memory (costing several hundred clock cycles twice).  Alternatively, if the memory in array position 1000000 was in position 1 instead, only one memory access would be required.}   
\item Shared memory is a very fast, small block of memory (16 kb on compute capability 1.3 and below, up to 48 kb for compute capability 2.0) which is accessible only within each block of threads.  
\item Texture memory is a cached global memory.  
\item Constant memory can only be assigned by the CPU and is a cached read-only memory.  
\end{enumerate}
Given the finite memory resources and speed, memory management is critical as most GPU algorithms are limited by their memory throughput \cite{CUDARef}.

\textbf{Computation:} Through the use of CUDA architecture and programming tools, the management and control of GPU computation and data parallelism is possible\footnote{CUDA provides a compiler and basic functions to perform computational tasks.  The CUDA tool-kit also includes a best practices guide which describes the advantages and limitations of GPU computing and how to get the best performance.}.  In CUDA, threads are organized into blocks which are executed on the same streaming multi-processor (SMP). Each GPU only has a finite number of SMPs and as a result can only computer a finite number of blocks at the same time. SMPs execute threads in groups (or warps).  Warps are chosen to be processed based on the availability of the requested memory resources. Thus, optimally choosing threads and threads per block can significantly enhance memory access and performance.

\textbf{Communication:} Any data dependency between threads requires special considerations.  Shared memory is the best method of dealing with any data dependencies.  This means that inter-thread communication is best handled within each block.  Communication between blocks can occur through a global sync between all GPU and CPU threads but is very inefficient.  Through the use of good parallel programming practices, the shared memory can be used efficiently to communicate between threads.  


We utilize CUDA programming to implement both the forward and inverse problems on GPU's. \emph{Our approach takes advantage of the GPU compute structure by designing an algorithm which minimizes data dependency between threads, maximizes the number of computations per global memory access, and minimizes CPU/GPU data communication.}  The optimization problem approach allows for minimal data dependency through the solving of many forward problems, that are independently solved on multiple threads on the GPU.

\subsection{Solving the forward problem}
Fast calculations of solutions to Eqn.~\ref{EB_ODE} are achieved through a CUDA based high-order ODE solver. The three key stages in solving the forward problem (FP) are: initialization, including memory set-up and modal function calculation; calculation of modal coefficients via high-order ODE solvers; and using modal functions and modal coefficients to calculate displacements.

\paragraph{Computational issues}     
  A Newton root solver and Simpson integration modules are used to solve for the modal resonance frequency ($\omega_i$) and normalization factor (to make the modal functions orthogonal), respectively.  Calculating the modal functions using the hyperbolic trigonometric form generally presented in texts \cite{Meirovitch2001} causes over-flow errors for higher modes. We recast the calculation to the following equivalent exponential form to enable accurate calculation without overflow:
\begin{eqnarray}
\Phi_i(x) = \frac{2 \sin (\beta_i x) - 2G \cos (\beta_i x) + (G-1) \exp^{\beta_i x} + (G+1) \exp^{- \beta_i x}}{2}, 
\end{eqnarray}
\begin{eqnarray}
G = \frac{2 \exp^{- \beta_i L} \sin (\beta_i L) - \exp^{-2 \beta_i L} + 1}{2 \exp^{- \beta_i L} \cos (\beta_i L) - \exp^{-2 \beta_i L} + 1}, 
\end{eqnarray}
where $\beta_i$ is the $i^{th}$ solution of $cos(\beta L) cosh(\beta L) + 1 = 0$ and $\omega_i = \beta^2_i \sqrt{\frac{EI}{\mu L^4}}$.  

Note that $\omega^2_i$ grows very quickly with increasing mode order $i$, making Eqn.~\ref{EB_ODE} a stiff ODE.  We utilize an explicit $4^{th}$ order Runge-Kutta (RK) method to solve the ODEs.\footnote{First order explicit Euler method was tested but failed to converge to a solution with practical time-step size of greater than $10^{-8}$. Second or third order methods have not been tested and could be a possible method of reducing calculation time if they converge.}   Explicit schemes were tested because they are efficient and the RK fourth order method converged in the range of time steps typical for AFM experiments. 

\begin{figure} [h]
\centering
\includegraphics[scale=0.5]{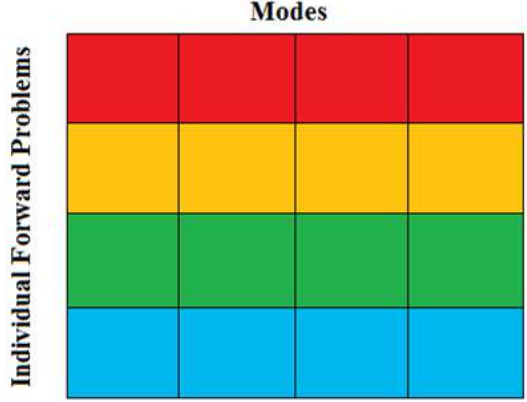}
\caption{Memory Allocation and Parallelization: A schematic of the forward problem parallel algorithm.  Squares represents an individual thread.  Each row represents an individual forward problem.  Each mode of each forward problem is given a thread.  Thus, each block will solve several forward problem solutions simultaneously.}
\label{Forward_Parallel_Alg}
\end{figure}

\paragraph{Memory allocation}
Solutions of the forward problem are obtained by calculating the modal functions ($\Phi_i$) and the modal coefficients ($\eta_i$). The modal functions remain invariant and thus only need to be calculated once.  We calculate and store the modal functions serially as part of the initialization process. Two parallel strategies are implemented to assist in solving Eq.~\ref{EB_ODE} for $\eta_i(t)$: {\it parallel prefix for force inner products and direct parallelism across modes}. Moreover, assuming that the force is variable separable (into spatial and temporal components), the spatial component of the force can be determined during initialization. This allows converting the computation of integrals involved on the RHS of Eqn.~\ref{EB_ODE} to a one time calculation. Pre-calculating force integrals leaves implementing a framework with parallelism across modes. Every modal coefficient ODE solve is handled by a unique thread. Thus, two dimensional blocks of threads are set-up as $(m, n_{fpb})$, where $m$ is the number of modes and $n_{fpb}$ is the number of forward problem solutions per block, as shown in Fig. \ref{Forward_Parallel_Alg}.  For example, on compute capability 1.3 GPUs, using eight modes, the current parallel framework can run up to 64 problems per block.

We next analyze the memory complexity of the framework. This elucidates the rational for deploying the various data structures in the available memory hierarchies. The major memory needs are as follows:
\begin{itemize}
\item The cantilever parameters, $E, I, \mu$. Since the cantilever parameters are assigned as part of the initialization process and require little memory, they are a good choice for constant memory. 
\item Parameters of the tip-sample interaction. This requires $4$ bytes for each parameter, thus requiring $8n_{fps}$ bytes for $n_{fps}$ simultaneous forward problem solutions using the visco-elastic model Eqn.~\ref{spring_damper}. Interaction parameters are left in the global memory since they have to be optimized in the inverse problem\footnote{The access speed of the interaction force parameters could potentially be improved by utilizing texture memory but has not been implemented in this work.}.
\item Modal coefficients $\eta_i$ require $4mn_{fps}$ bytes, where $m$ is the number of modes used. The output deflection points require $n_{dp}n_{fps}$ bytes, where $n_{dp}$ is the number of deflection points computed. Usually deflections at three points on the AFM cantilever are measured ($n_{dp} = 3$). The modal coefficients and deflection points are stored in shared memory because of the constant updating during the forward solve.  
\end{itemize}

\paragraph{Complexity analysis}
  For the forward problem, the main calculations affecting runtime complexity are the force integral calculations and the ODE solve. The runtime complexity for the serial forward problem is $\mathcal{O}([n_t n_x + n_t]m)$, where $m$ is the number of modes, $n_t$ is the number of time steps and $n_x$ is the number of spatial points used to compute the force integral.  However, assuming variable separation of the forces reduces serial runtime complexity to $\mathcal{O}(mn_t)$. By deploying across $m$ threads on a GPU, the parallel runtime complexity is $\mathcal{O}(n_t)$.  

\begin{figure} [ht]
\centering
\includegraphics[scale=0.2]{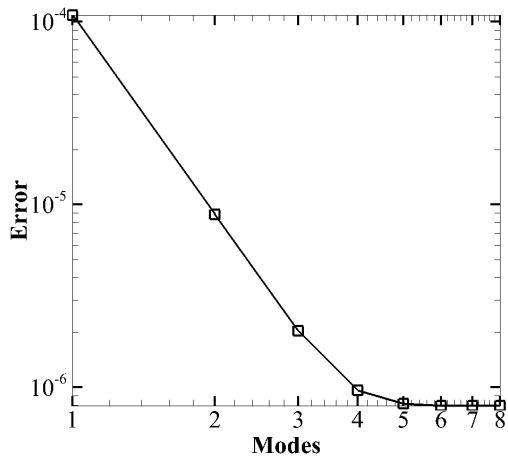}
\caption{Convergence compared to 8 modes for simple sinusoidal base movement}
\label{SerialConvergence}
\end{figure}
  
\subsection{Speedup characteristics: Comparing CPU vs GPU implementation}
In addition to the GPU based implementation discussed in the preceding section, we also implemented a CPU based version of the ODE solver for comparison. The validity of the results of both implementations are ensured by comparing with analytical solutions that are obtained using a sinusoidal base movement and constant forces~\cite{Meirovitch2001}(see Appendix A).  
\begin{figure} [ht]
\centering
\mbox{
\subfigure[]
{\includegraphics[scale=0.2]{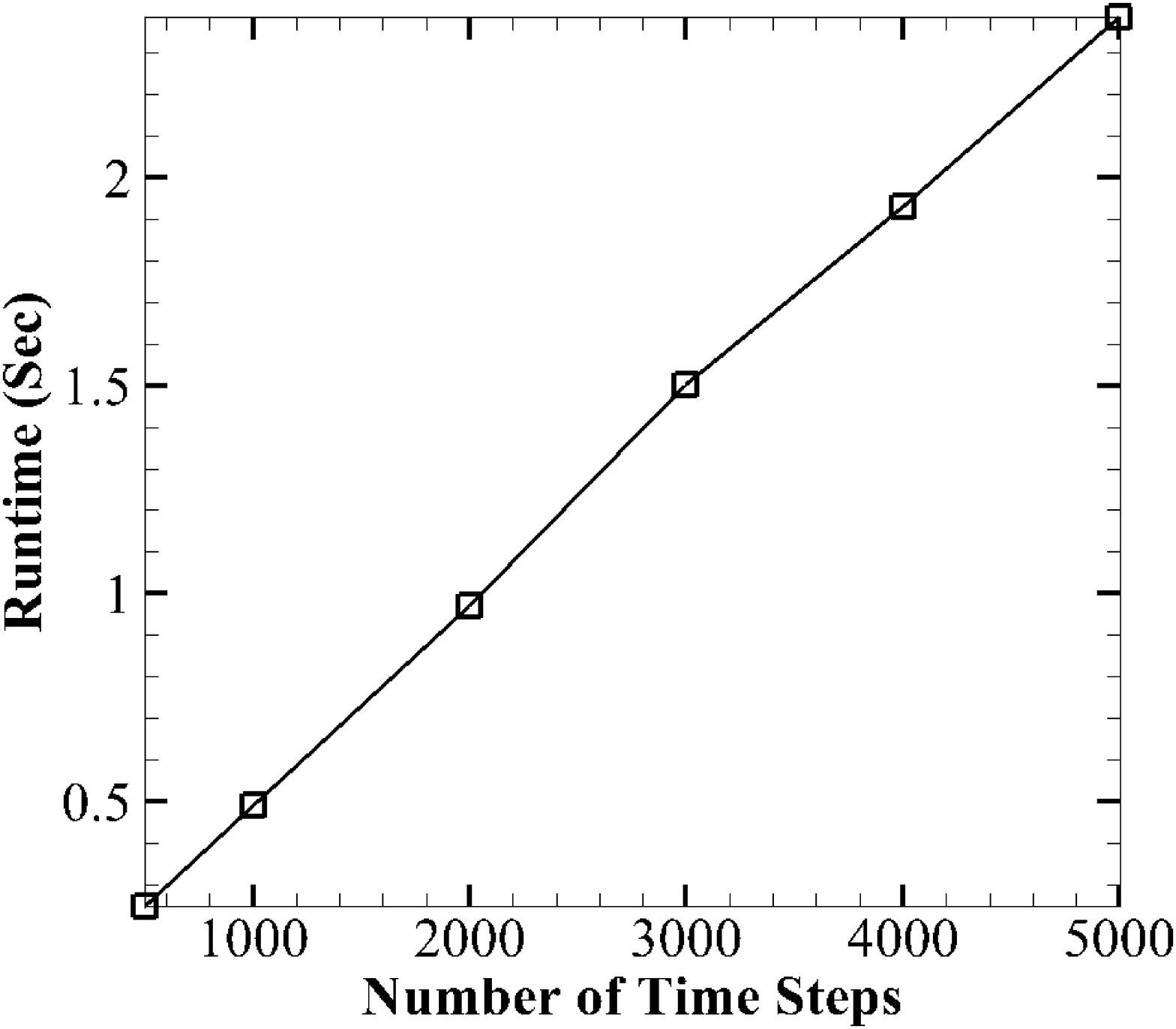}
\label{SerialNumTimePointsvRuntime}}
\quad
\subfigure[]
{\includegraphics[scale=0.2]{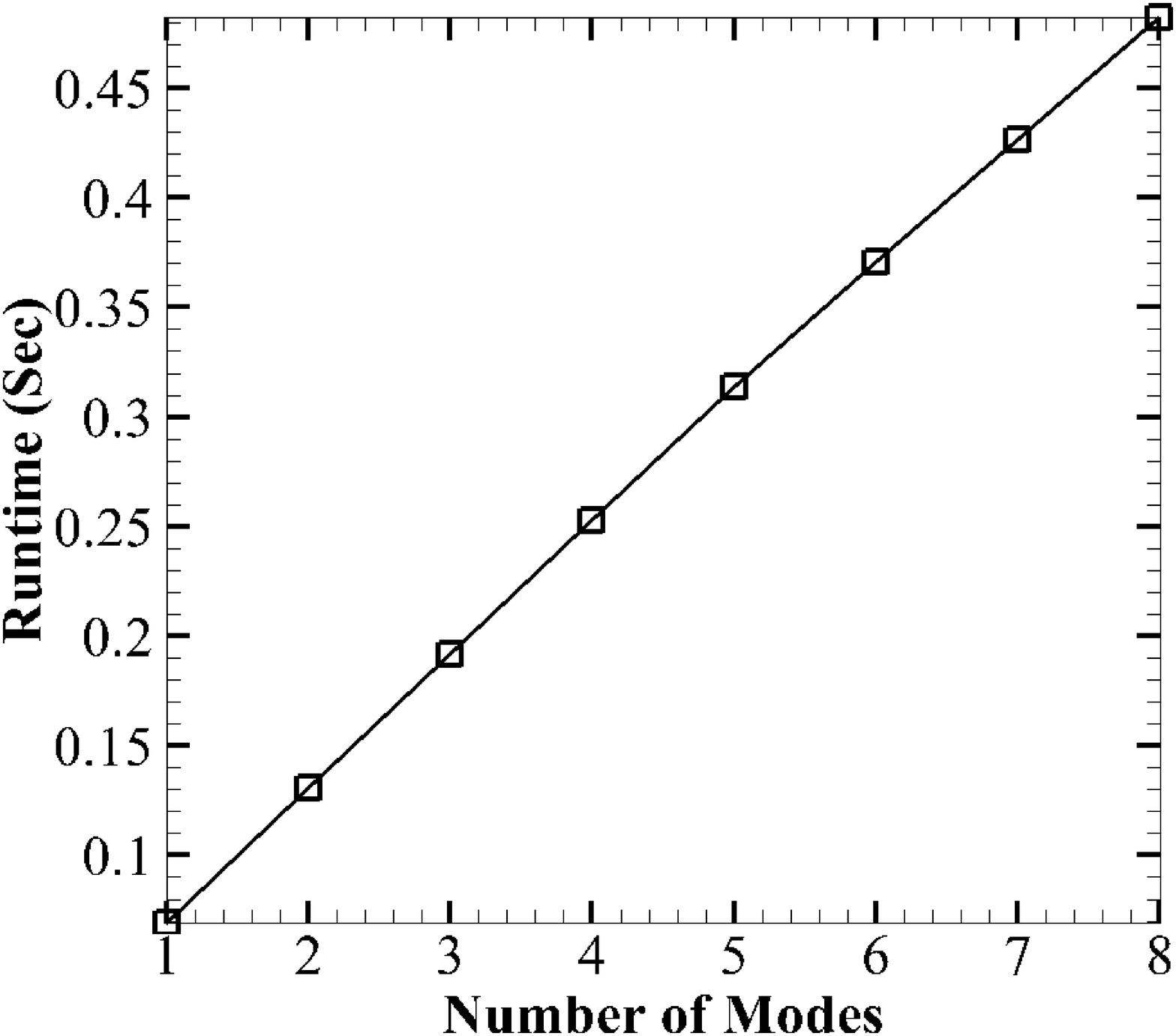}
\label{SerialModesvRuntime}}
}\caption{(a) Number of time points vs. runtime in seconds for CPU framework; (b) Number of modes vs. runtime in seconds for CPU based framework}
\end{figure}

We discuss runtime trends and accuracy details in this subsection. Each forward problem is run for 1000 time steps (unless otherwise stated) using time-step, $\Delta t = 10^{-7}$. A maximum of eight modes are used. Eight modes can satisfactorily track the cantilever evolution (with an error of $3.310^{-6}$).  Error is defined as $\frac{|V_t - V_c|_{L2}}{|V_t|_{L2}}$, where $V_t$ is the true value and $V_c$ is computed.  In testing convergence, the L2 error of various modes are compared to the solution obtained using eight modes.  Fig. \ref{SerialConvergence} shows the plot of error versus number of modes. For this simple case, four modes are sufficient for resolving a sinusoidal deflection.   

While considering runtime complexity three parameters are most dominant; number of modes, number of points used to describe modal functions, and number of time points. We analyze all three of them independently, first for the CPU based implementation and subsequently for the GPU based implementation. Runtime as a function of the number of modes is shown in Fig. \ref{SerialModesvRuntime} with the number of time points fixed at 1000 (note that 8 modes take 0.482 seconds). Fig.~\ref{SerialNumTimePointsvRuntime} plots runtime as a function of the number of time points. When seeking real-time inversions, runtime must be of the order of few hundred milliseconds.  These runtime analyses show that calculation times using CPUs are too slow to solve the inverse problem in real time. 

\begin{figure} [ht]
\centering
\includegraphics[scale=0.2]{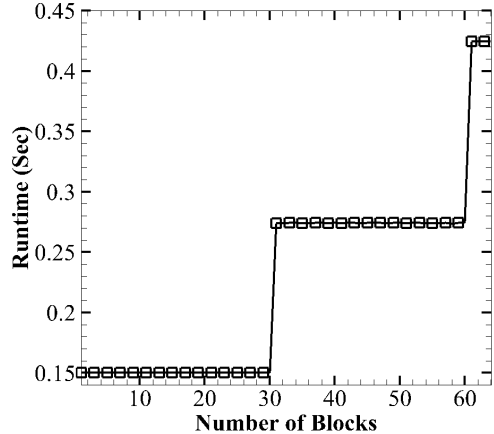}
\caption{Number of blocks vs. runtime in seconds}
\label{BlockvRuntime}
\end{figure}

In contrast to the CPU results, the GPU based results are very promising. With the goal of solving several hundreds of forward problems to solve one inverse problem, we utilize two metrics to illustrate the capabilities of the GPU based framework: (a) runtime of individual forward solves; (b) the number of forward solves which can be calculated in parallel. Both metrics depend on multiple factors: number of time steps ($n_t$), number of modes ($m$), number of solutions per block ($n_{fpb}$), and total number of forward problem solutions ($n_{fps}$). A Nvidia Quadro FX 5800 GPU is used which limits the number of blocks that can run in parallel to 30 (2 blocks per streaming multiprocessor(SMP), 15 SMP's)\footnote{For more details on the system used for testing and implementation, see appendix C.}.  Beyond this, with all SMPs filled, the blocks have to wait for an open SMP. This can be clearly seen in Fig. \ref{BlockvRuntime} where the runtime jumps after 30 and 60 blocks.  
\begin{figure} [ht]
\centering
\mbox{
\subfigure[]{
\includegraphics[scale=0.2]{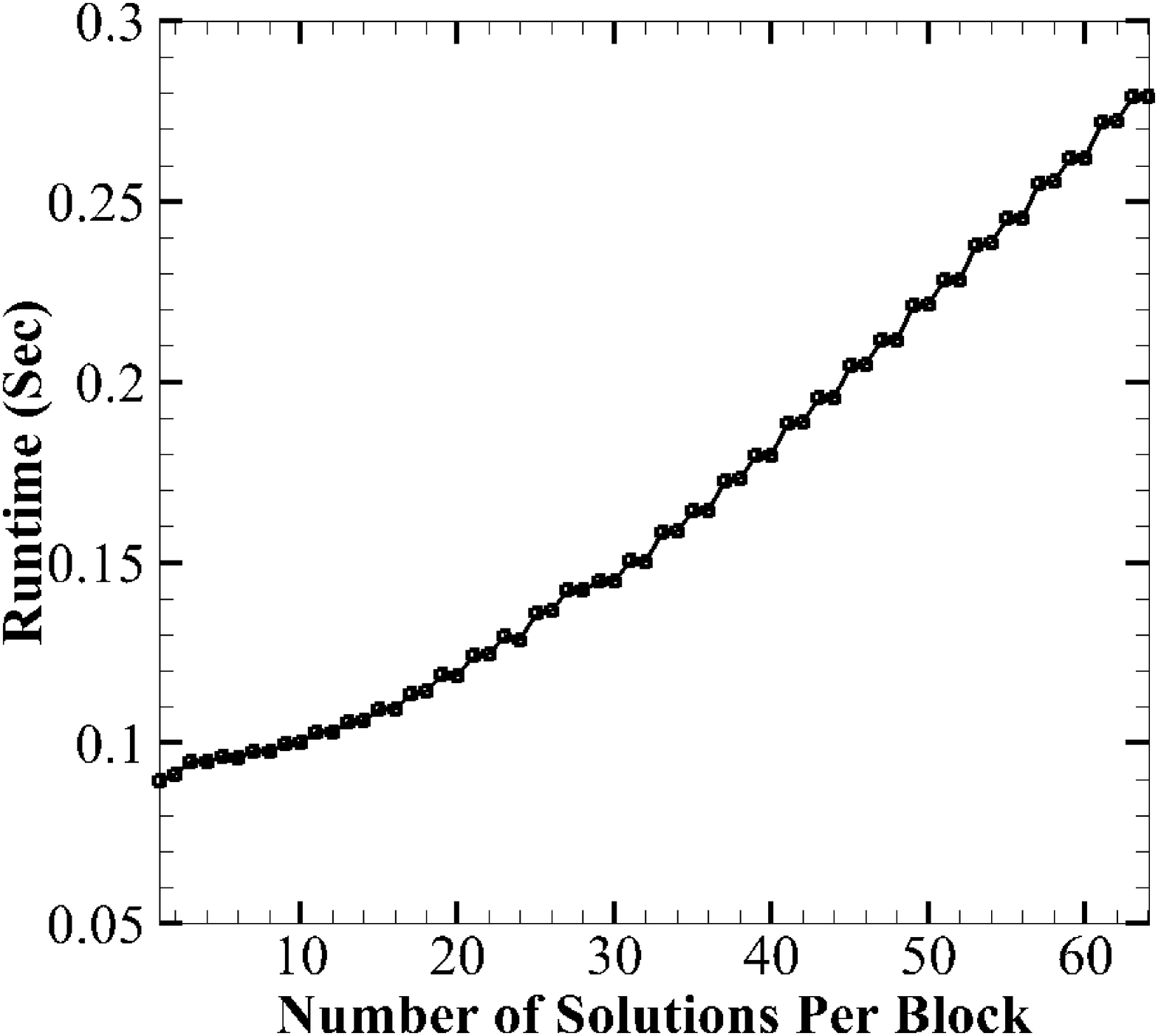}
\label{SPBvRuntime}}
\quad
\subfigure[]{
\includegraphics[scale=0.2]{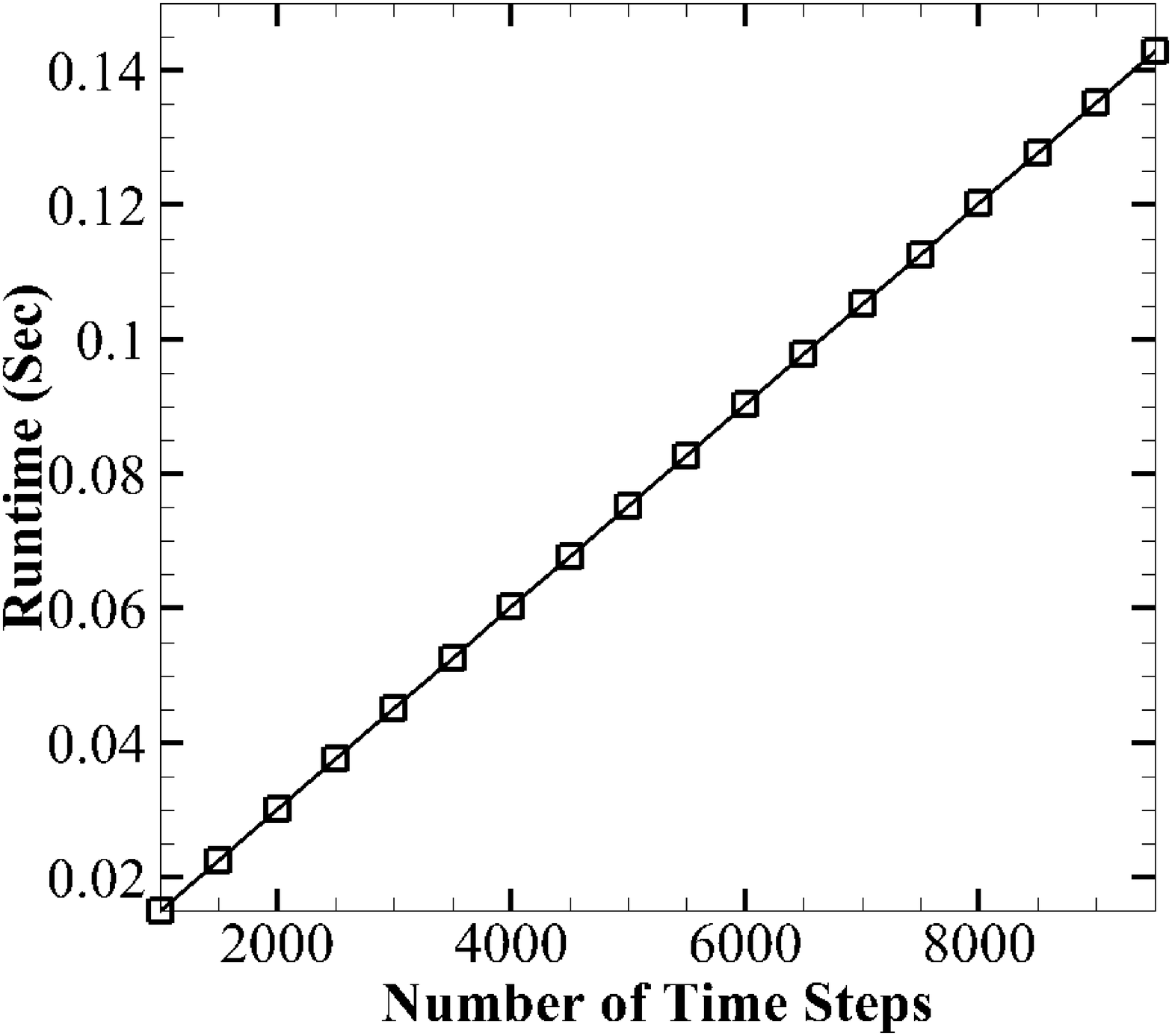}
\label{NumTimePointsvRuntime}}
}\caption{(a) Number of solutions per block vs. runtime in seconds; (b) Number of time points vs. runtime in seconds}
\end{figure}
Fixing the number of blocks ($\frac{n_{fps}}{n_{fpb}}$) to 30 and modes ($m$) to 8, runtime is analyzed by varying the number of time steps and number of solutions per block.  Fig. \ref{SPBvRuntime} show the effect of $n_{fpb}$ on runtime.  As the number of solutions per block increases, the runtime increases in a non-linear way. 32 solutions per block most efficiency utilize GPU resources giving forward solves with 100 millisecond runtime. Furthermore, as $n_{fpb}$ increases beyond 32, the solutions per second gain is small while 32 solutions only requires half of the shared memory.  Fig. \ref{NumTimePointsvRuntime} shows that increasing $n_t$ causes a linear increase in runtime (as predicted by the complexity analysis in the previous section).  

By appropriately choosing the number of solutions per block to ensure proper memory allocation, the runtime for a single forward solve on the GPU took $0.0151$ seconds. {\it This is a speed-up of 32 over the CPU framework.} More importantly, the GPU based framework can execute several forwards solves simultaneously. {\bf 960 forward solves are computed in 0.0151 seconds, translating to an effective speed-up of 30000}.  Effective speed-up is the direct comparison between our CPU (single-core) performance and GPU performance.  

\subsection{Inverse problem algorithm}
The choice of the optimization algorithm is driven by the following constraints:
\begin{enumerate}
\item The GPU based forward solver implementation is able to compute several forward solution in parallel.
\item The parametrization of the cost function $\mathcal{J}$ may be high dimensional. 
\item Furthermore, the landscape of $\mathcal{J}$ may be non-smooth, necessitating a gradient free method.
\item The existence of multiple local minima that have to be discarded.
\end{enumerate}
A gradient-free, global search algorithm that satisfies these constraints is the particle swarm optimization (PSO) \cite{PSO_Orignal}. PSO finds the global minima by starting with a large population of candidate solutions (or particles), and moving these particles around in the search-space according to certain rules over the particle position and velocity. Each particle is influenced by its local best known position and the best known positions in the search-space, which are updated at every iteration as better positions are found by other particles. Generally, particle locations and velocities are chosen using a uniform distribution in the search space~\cite{PSO2002,PSO2006}. The update of velocity uses the following equation:
\begin{eqnarray}
v_i = w v_{i-1} + c_1 r_1 (b_l - x_{i-1}) + c_2 r_2 (b_g - x_{i-1}),
\end{eqnarray}
where $v_i$ is the velocity at iteration $i$, $w$, $c_1$, and $c_2$ are weighting factors, $r_1$ and $r_2$ are random numbers, $b_l$ and $b_g$ are the local and global bests respectfully, and $x_i$ is the position of the particle at iteration $i$. Recent theoretical results suggest that an appropriate choice of the the parameters guarantee convergence~\cite{PSO2002,PSO2006}. We utilize an optimized GPU based implementation of the PSO algorithm~\cite{CUDAPSO}.


\section{Results}



This section discusses a hierarchy of increasingly complex inverse problems. We start with the simpler problem of extracting the base vibration characteristics given tip deflections in Section.~\ref{P1}. This problem also explores the choice of the search space parameters and their effect on runtime. The next two subsections deal with real time inversions of elastic and visco-elastic tip-sample interactions, respectively.

\subsection{Base vibration force calculation} \label{P1}

'Experimental' tip deflection data was computed by forcing the base to vibrate to a simple sinusoidal driven signal:
\begin{eqnarray}
y(t) = a~ sin(2 \pi f t),
\end{eqnarray}
where $a$ is the amplitude and $f$ is the frequency of cantilever base vibration.  The 'experimental' tip deflection was obtained by setting $a~=~2nm$ and $f = 25600~ Hz$. This 'experimental' tip deflection -- subsequently used to drive the inverse problem -- is obtained using the CPU based serial framework, thus resolving the issue of inverse crime \cite{inverse_crime}.  

\begin{figure} [ht]
\centering
\mbox{
\subfigure[]{\includegraphics[scale=0.2]{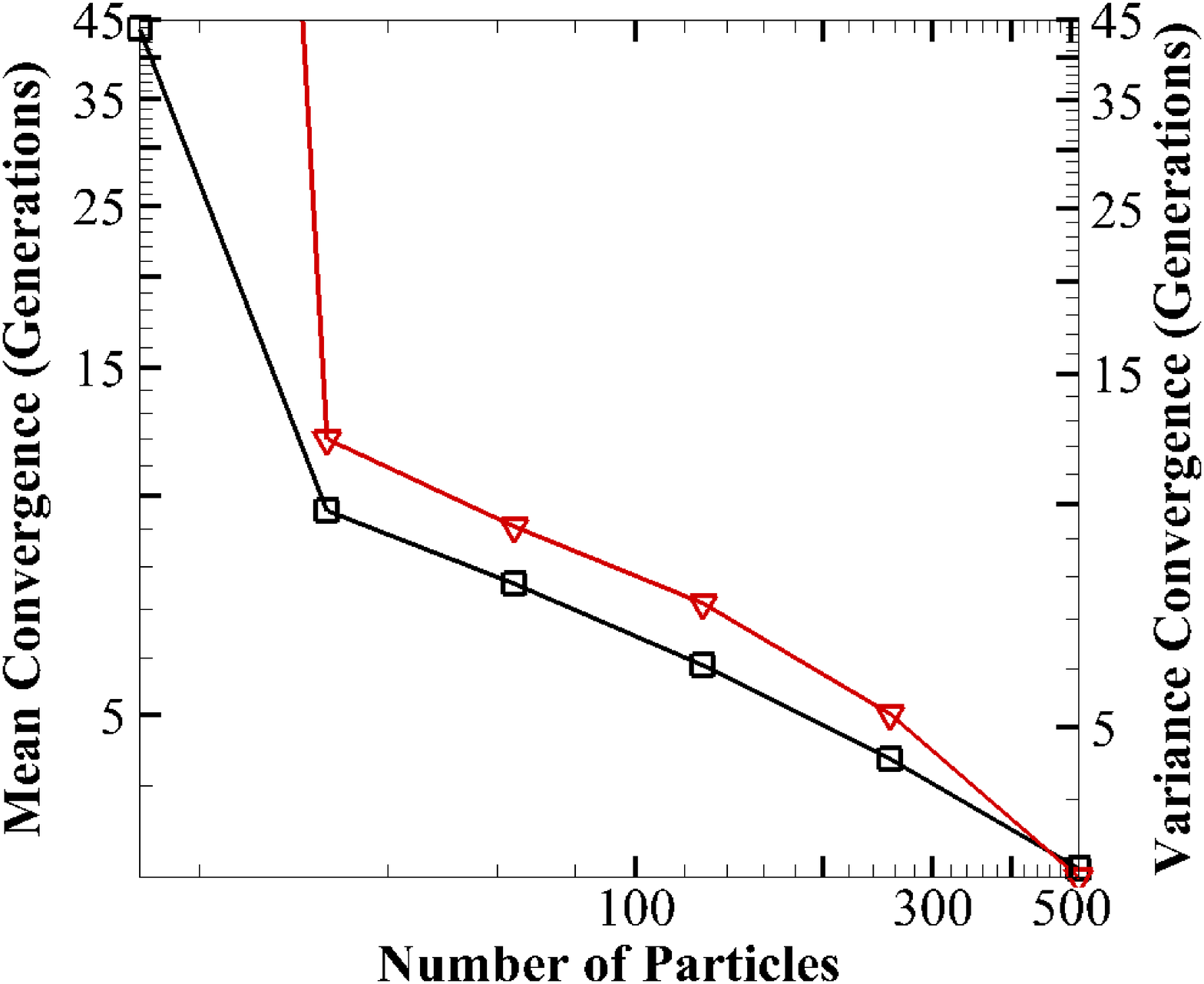}
\label{ParVGen1D}}
\quad
\subfigure[]{\includegraphics[scale=0.2]{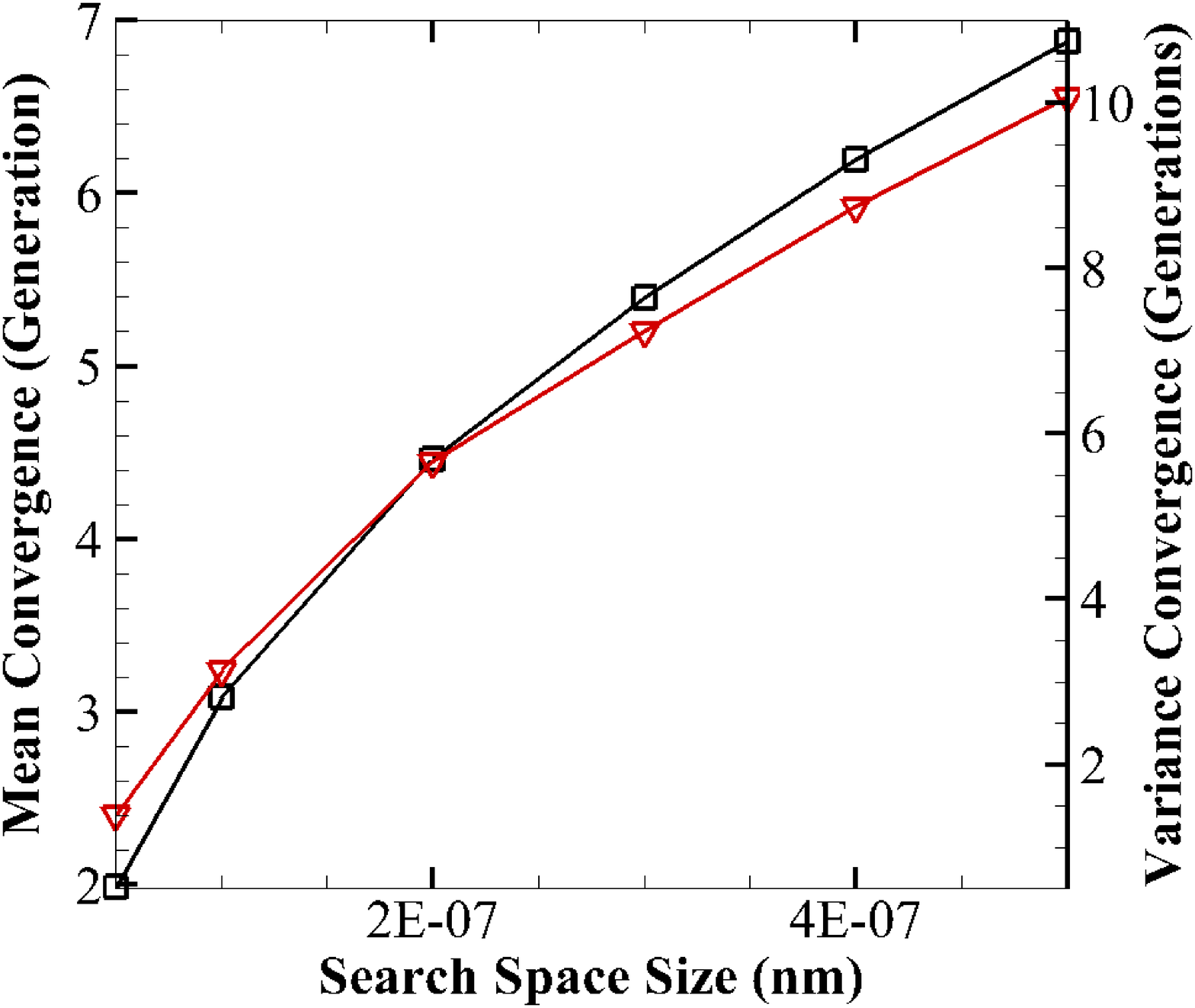}
\label{SearchSpaceVGen1D}}
}\caption{(a) Number of particles effect on mean($\square$) and variance($\nabla$) of generations to convergence with 1D search space; (b) Search Space effect on mean($\square$) and variance($\nabla$) of generations to convergence with 1D search space}
\end{figure}

We analyze the performance of the inversion framework by starting with a one dimensional search space for the amplitude of base vibrations and fixing $f = 25600~ Hz$.  We provide physically meaningful bounds on the amplitude, $[0~ nm, 100~nm]$ and set $n_t = 1000$. The number of time steps was chosen to provide a sufficient number of data points per oscillation period given the typical time-step used.  Fig. \ref{ParVGen1D} shows that the mean and variance of the number of generation for convergence shrinks as the number of particles used in the PSO increases.  Faster convergence is expected since the density of the particles in the search space is increasing.  Given that the framework produces a generation about every 16 milliseconds, {\it the mean convergence time for the 1D case is just over 48 milliseconds for 512 particles}.    

We next investigate the affect of search space size on the performance of the inversion framework, as shown in Fig. \ref{SearchSpaceVGen1D}. In general, as the particle density in the search space decreases, the mean number of generations required for convergence increases.  For the case of the largest search space ($[0 ~nm, 500~nm]$) the mean number of generation of 6.8767 corresponds to a runtime of around 112 milliseconds.  \emph{This suggests that a tight bound on the unknown force parameters can significantly decrease inversion times.}

\begin{figure} [ht]
\centering
\mbox{
\subfigure[]{\includegraphics[scale=0.275]{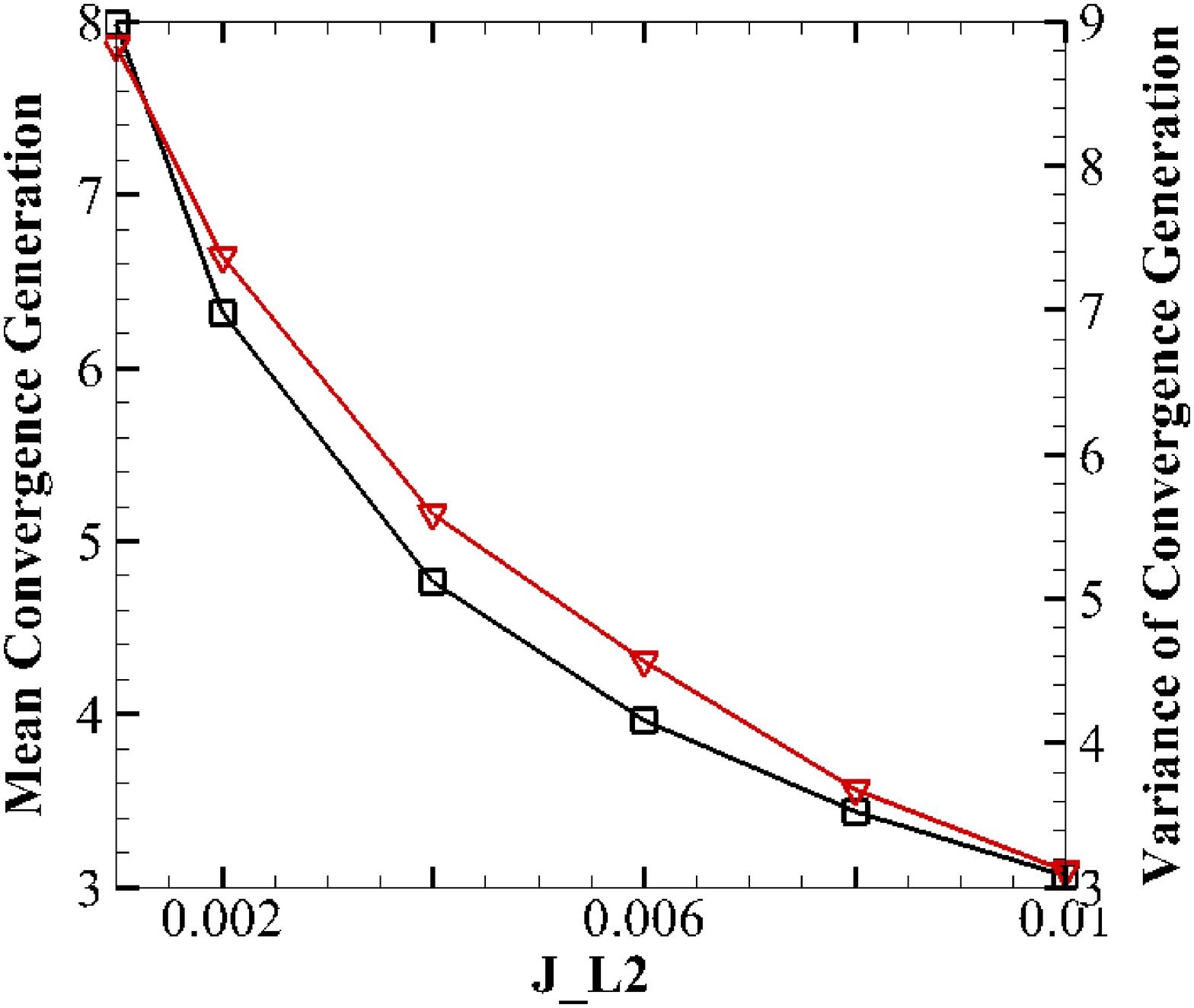}
\label{AccVGen1D}}
\quad
\subfigure[]{\includegraphics[scale=0.2]{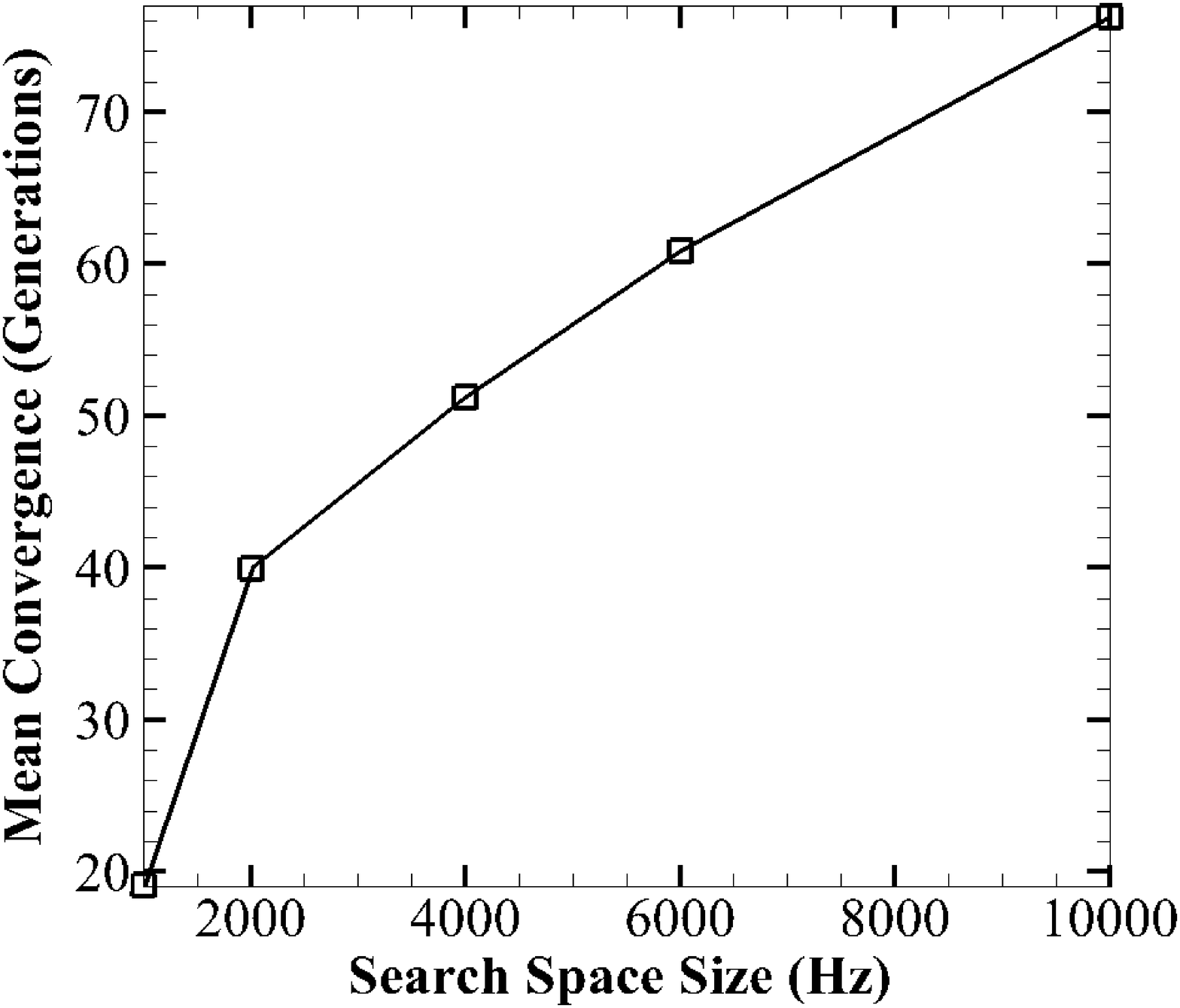}
\label{SearchSpaceVGen2D}}
}\caption{(a) $\mathcal{J}_{L2}$ target effect on mean($\square$) and variance($\nabla$) of generations to convergence with 1D search space; (b) Search Space effect on mean($\square$) of generations to convergence with 2D search space}
\end{figure}

The affect of convergence cutoff on convergence is shown in Fig. \ref{AccVGen1D}.  The general trend of increasing accuracy requirements raises the mean number of generations required for convergence.  To achieve an error of 0.001 requires an average of 7.99 generations (taking 128 milliseconds) verses the 48 milliseconds required for achieving a relative error of 0.01. The effect of $n_t$ on mean generations was also tested.  The number of time steps was varied from 1,000 to 10,000 and did not show any effect on the result.

We next tested the inversion framework by using a 2D search space (making both $a$ and $f$ unknown). Fig. \ref{SearchSpaceVGen2D} shows a similar trend to the 1D case where the mean number of generations required for convergence increases as the search space increases.  With a search space of 1000 Hz, 19.0458 generation on average is required for convergence resulting in a compute time of about 320 milliseconds.  

To explore frequency dependent properties, the AFM can be driven by a chirp base vibration. This linearly varies the frequency of base oscillation with time:
\begin{eqnarray}
y(t) = a~ sin(2 \pi [f + g t]t),
\end{eqnarray}
where $g$ is defined as the frequency gain parameter. We utilize the inversion framework to extract the parameters of this chirp signal. Matching a chirp base movement requires exploring a 3D search space for $(a,f,g)$. The 'experimental' tip deflection was obtained by setting $a=~2~nm$, $f = 10,000~Hz$ and $g = 20,000,000~ \frac{Hz}{Sec}$. Using a search space of $[0~nm, 100~nm]\times [7,500~Hz, 12,500~Hz] \times [19,000,000~\frac{Hz}{Sec}, 21,000,000~\frac{Hz}{Sec}]$, the framework extracts correct values of $(a,f,g)$ in 14.9 generations corresponding to a runtime of about 240 milliseconds. 

\subsection{Inverting elastic properties}
\begin{figure} [ht]
\centering
\mbox{
\subfigure[]{
\includegraphics[scale=0.275]{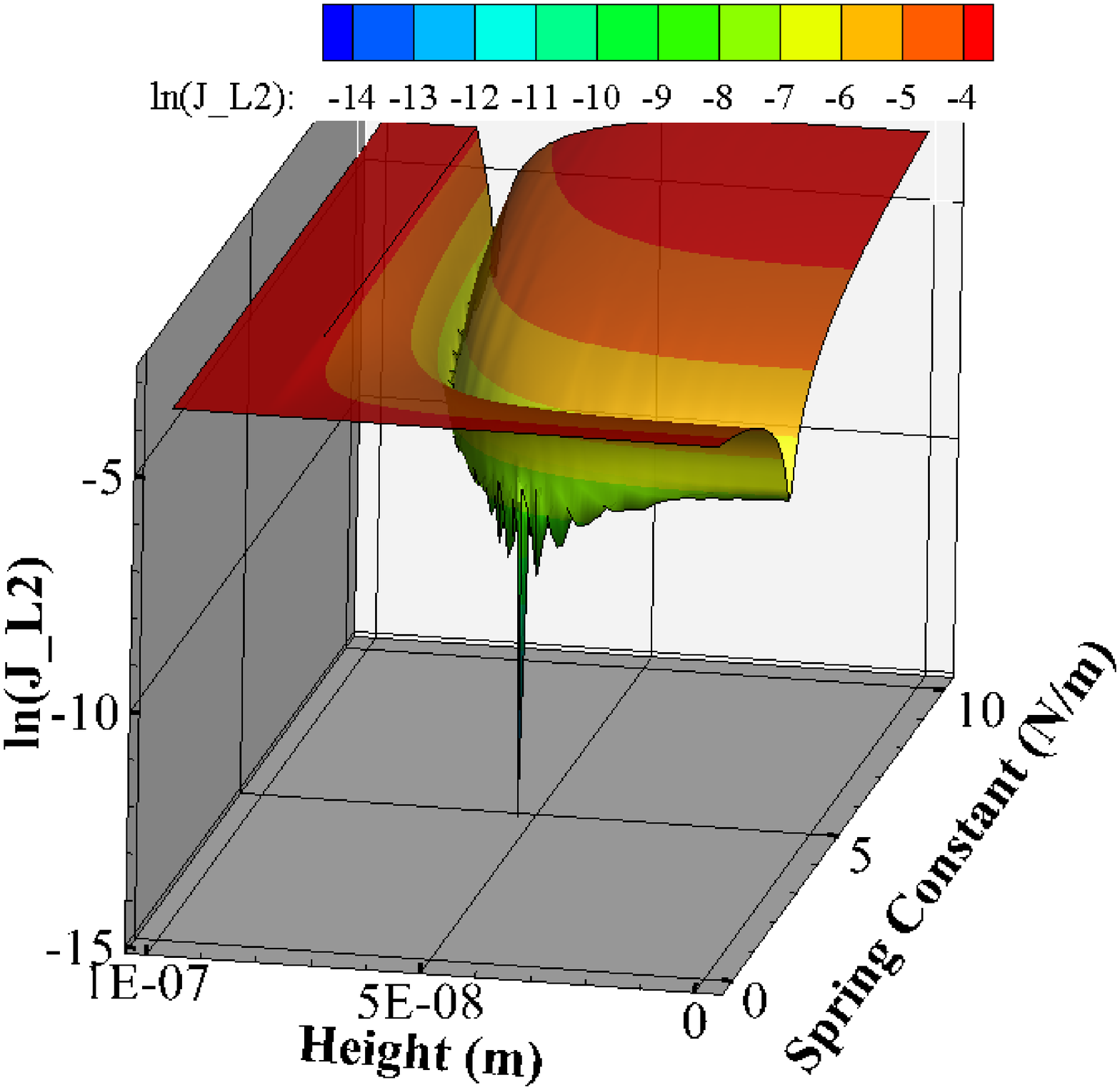}
\label{SearchSpaceSpring}}
\quad
\subfigure[]{\includegraphics[scale=0.275]{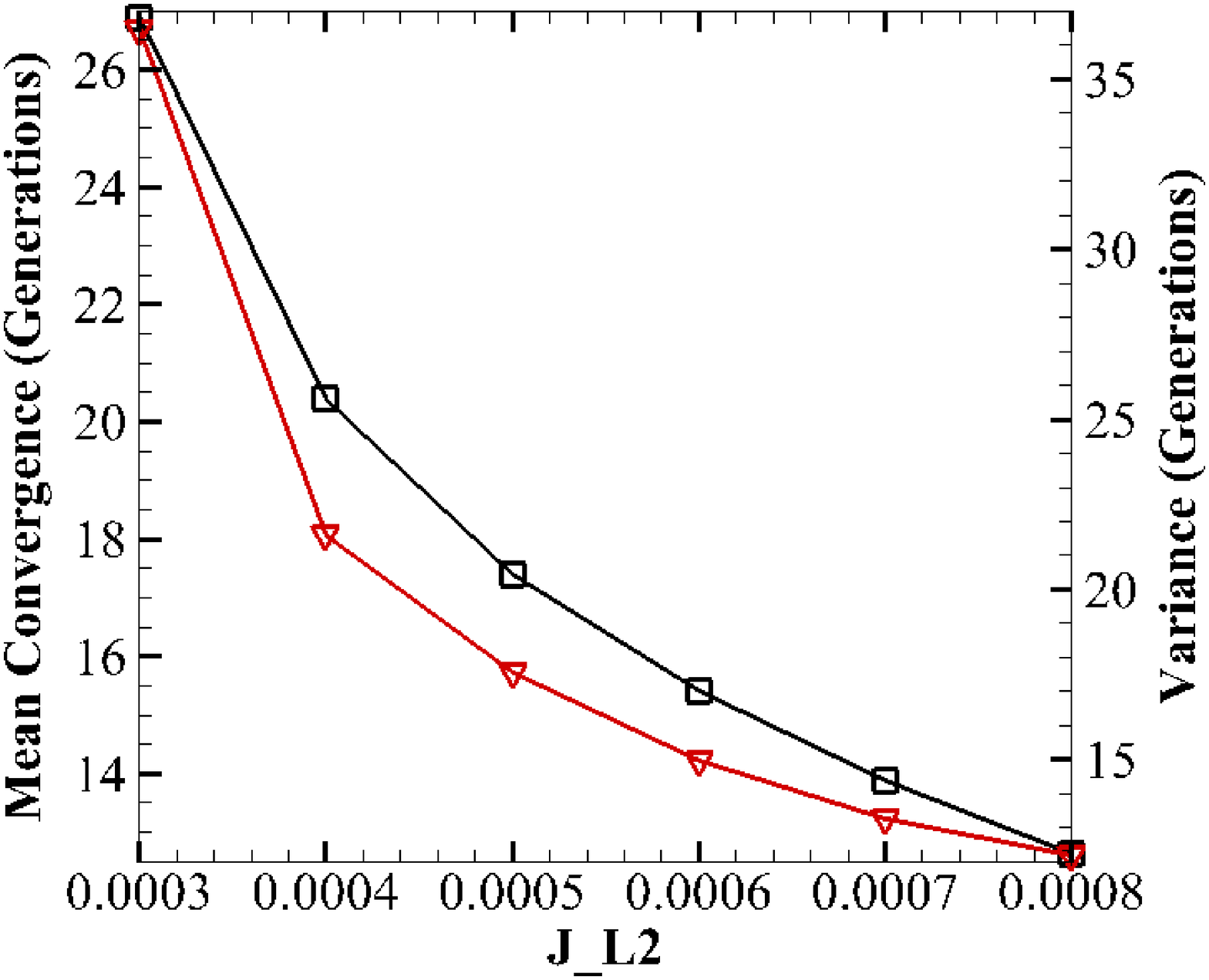}
\label{SpringConverge}}
}\caption{(a) $\mathcal{J}_{L2}$ Space of spring and cantilever-sample separation; (b) $\mathcal{J}_{L2}$ requirement effect on mean($\square$) of generations to convergence with spring tip-sample search space}
\end{figure}
A simple yet extensively used model for tip-sample interaction is one that assumes an elastic response of the soft sample. Extracting spatial variation in elasticity is important for a variety of applications in addition to non-destructive scanning of the sample (e.g., tumors have higher stiffness that normal cells; as collagen dehydrates there is a change in elasticity~\cite{LinH99,Feninat2001}). 

We utilize a Hooke's law based parametrization for the elastic response of the sample. The modelling of the tip-sample interaction as a spring is in-line with most AFM experimental force calculation models. This model has the following form:
\begin{eqnarray}
f(t) = -k (u - h)
\end{eqnarray}
where $u-h$ is the distance the cantilever tip has pressed into the sample and k is the tip-sample spring constant. With the base movement parameters known, only two tip-sample force parameters are unknown, the cantilever-sample separation and sample spring constant.  Cantilever-sample separation is defined as distance from the cantilever's neutral axis to the sample.  The 'experimental' tip deflection was created using the following parameters: $h = 50~nm,$ and $k = 4 N/m$, where $h$ is the cantilever-sample separation and $k$ is the sample spring constant.  These parameters were chosen to mimic results found in \cite{Raman08}.  

Calculating the force parameters for a spring sample is computationally complex due to the non-convex nature of the phase space as shown in Fig. \ref{SearchSpaceSpring}, and results in more calculation time being required shown in Fig. \ref{SpringConverge}.  Fig. \ref{SearchSpaceSpring} shows that several combinations of $k$ and $h$ produce similar values, creating a symmetrical valley near the basin of the global minima. The resulting valley of similar combinations is expected because both $k$ and $h$ only affect the amplitude of vibration and not the phase. Using a cutoff convergence threshold of $\mathcal{J}_{L2} = 0.0005$ results in a 5\% error in calculating $k$ and $h$ resulting in a mean convergence of about 18 generations, resulting in an average runtime of 288 milliseconds.

\begin{figure} [ht]
\centering
\mbox{
\subfigure[]{\includegraphics[scale=0.275]{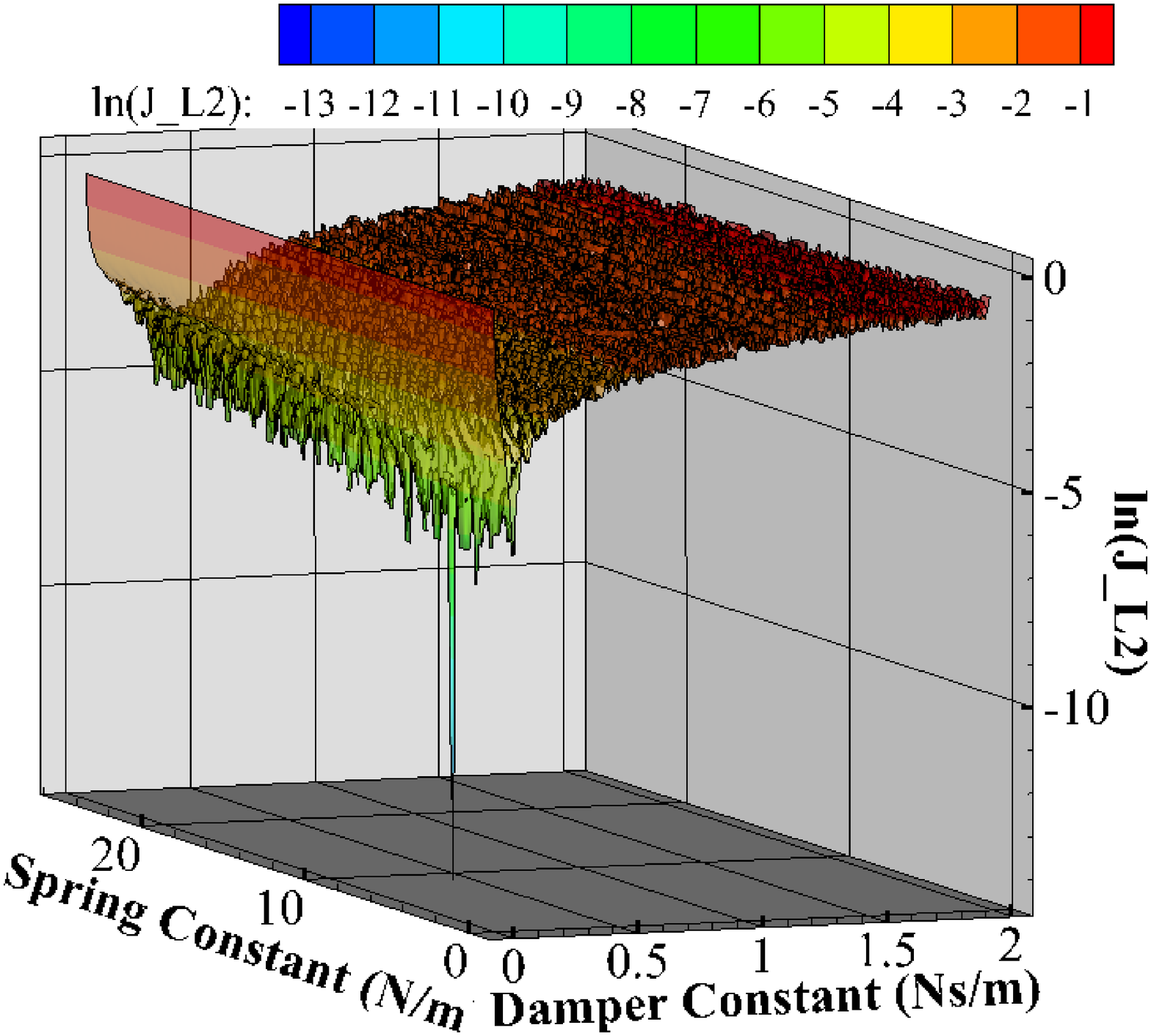}
\label{SearchSpaceSpringDamp}}
\quad
\subfigure[]{\includegraphics[scale=0.275]{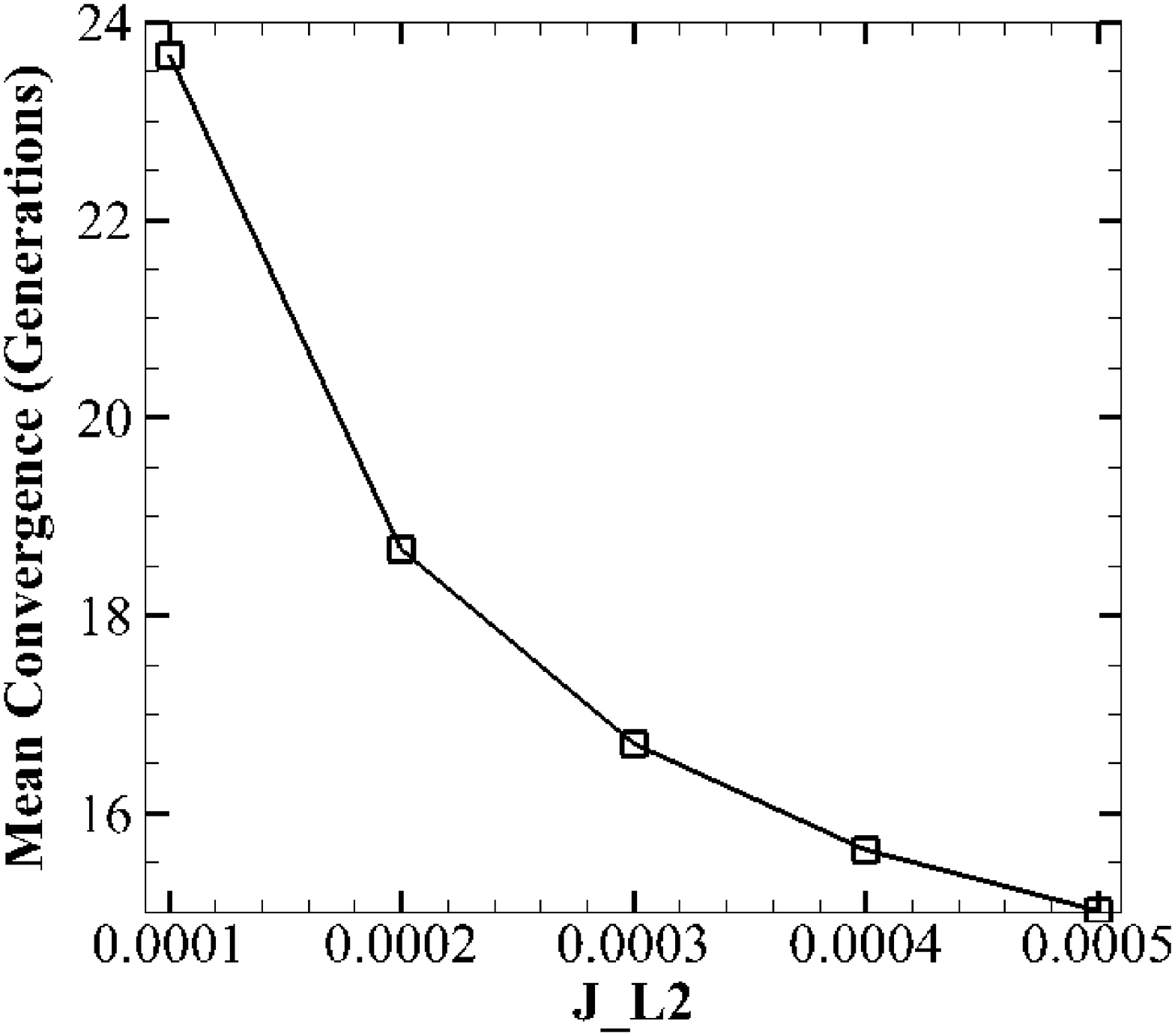}
\label{SpringDampConverge}}
}\caption{(a) $\mathcal{J}_{L2}$ Space of spring and damper; (b) $\mathcal{J}_{L2}$ requirement effect on mean($\square$) of generations to convergence with spring tip-sample search space}
\end{figure}

\subsection{Viscoelastic tip-sample interaction}
We increase the complexity of the tip-sample interaction parametrization by next assuming a visco-elastic response of the sample. Understanding the visco-elastic variations is particularly important in understanding aspects of polymer physics\cite{Garcia06}. A Kelvin-Voigt parametrization is extensively used to model visco-elastic tip-sample interactions \cite{Raman08}. This is essentially a spring damper system:
\begin{eqnarray}
f(x,t) = -k (u - h) -\zeta_s \dot{u}
\end{eqnarray}
where $\zeta_s$ is the viscous damping coefficient. We assume that the sample separation for the AFM cantilever neutral axis is known. This inversion consists of estimating two material properties.
The 'experimental' tip deflections were obtained by using the following parameters: $h = 50~nm,$, $k = 4 N/m$, and  $\zeta_s = 0.1 \frac{Ns}{m}$. Using a search space of $[0~ N/m, 25~N/m] \times [0~ \frac{Ns}{m}, 2~ \frac{Ns}{m}]$ the convergence profile with increasing number of generations of the PSO scheme is shown in Fig. \ref{SpringDampConverge}.  

The phase space for inverting the visco-elastic tip-sample interaction is highly corrugated and has multiple local minima as shown in Fig. \ref{SearchSpaceSpringDamp}.  The shape of search space clearly demonstrates the need for gradient free optimization methods.  The large number of local minima is due to the phase change caused by the damping force.  The average runtime for inversion (with 16 generations) was 256 milliseconds.  

\section{Conclusions}

The atomic force microscope (AFM) is a versatile, high-resolution scanning tool used to characterize topography and material properties of a large variety of specimens. Its applicability to characterize soft specimens like tissue and gels is currently constrained by the inability to appropriately control tip-sample interaction forces. A major bottleneck to control tip-sample interaction is the ability to extract these tip-sample forces in real time from the deflection signal. This paper illustrates a first approach to a near real-time framework for tip-sample force inversion. We utilize the hardware advantages and parallel capabilities of GPUs to develop a fast inversion strategy. A fast, parallel forward solver is developed that shows a 30000 fold speed-up over a comparable CPU implementation, resulting in milli-second calculation times. Posing the inverse problem as an unconstrained optimization problem allows us to integrate a GPU based gradient-free global Particle Swarm Optimization framework with the forward solver. We illustrated the framework on three classes of tip-sample interaction inversions. Each of these inversions is performed in sub-second timings showing potential for on-line integration with the AFM. Extensions of this work include (a) integrating with AFM hardware for deployment; (b) investigating more complex tip-sample parametrization that account for adhesion and Van der walls forces; and (c) perform nano-composition mapping of soft tissue.

We envision that the developments presented here will find applicability in a broad spectrum of
areas that require soft-imaging as well as nano-composition imaging. These range from nano-material synthesis
and design for emerging applications such as solar cell and micro-scale energy devices; interrogating
cellular and sub-cellular interactions; investigating the optoelectronic properties of organic electronic devices to understanding reaction initiation in High Energy Density Materials (HEDM). In particular, this framework should open the door to interrogate various bio-chemical interaction forces (hydrophobic interaction force) of live cell for understanding fundamental biology mechanisms such as the cell fusion process.

\section{Acknowledgements}
B.G. was supported in part by NSF PHY-0941576, and NSF CCF-0917202. The authors would like to thank the high-performance (GPU) computing resources at ISU.

\appendix

\section{Dynamic cantilever model data}
\begin{center}
\begin{tabular}{|c|c|}
\hline L2 error & $0.00333531$ \\ 
\hline Length & $525$ microns \\ 
\hline Elasticity & $1.76 * 10^{11}$ Pa \\ 
\hline Width & $35$ microns \\ 
\hline Height & $5$ microns \\ 
\hline Inertia & $3.64583$ kg $m^2$ \\ 
\hline Mass per unit length & $4.0775 * 10^{-7}$ $\frac{kg}{m}$ \\ 
\hline Start Time & $0$ Sec \\ 
\hline End Time & $0.004$ Sec \\ 
\hline Time step & $4 * 10^{-8}$ Sec \\ 
\hline Number of modes & 8 \\ 
\hline Number of modal function points & 501 \\ 
\hline Amplitude & $5$ nm \\ 
\hline Frequency & $25500$ Hz \\ 
\hline First natural frequency & $25468.9$ Hz \\ 
\hline Steady State Deflection Amplitude & $3.2$ microns \\ 
\hline 
\end{tabular} 
\end{center}

\section{Modal analysis}
The deflection of the cantilever then is represented as:
\begin{eqnarray}
u(x,t) = \sum_{i=1}^N \eta_i (t) \Phi_i (x),
\end{eqnarray}
where $\Phi_i (x)$ represents a set of orthogonal modal shape functions which are computed by solving the homogeneous eigenvalue problem  \cite{Meirovitch2001} and $\eta_i (t)$ are corresponding modal coefficients.  In order to determine the modal coefficients, we define an inner product as:
\begin{eqnarray}
\langle f, g \rangle = \int_L f g dx,
\end{eqnarray}
where $f$ and $g$ are functions.  

\section{Hardware}
Testing and implementation of the proposed framework occurred on a GPU mini-cluster consisting of two Dell Precision T7500 workstations.  Each node is equipped with 12 GB of DDR3 RAM, 500 GB of 10K RPM hard drives and 2 Intel Xeon 2 GHz quad-core CPUs. The main computational power comes from 4 GPUs donated to us by NVIDIA: each node is accelerated with 2 NVIDIA QUADRO FX 5800. One such card provides 240 cores, 4 GB of RAM with 102 GB/s bandwidth and is CUDA compatible (with 1.3 compute capability). The nodes are connected via dedicated Gbit Ethernet.







\end{document}